# All Dielectric Metamaterial Loaded Tunable Plasmonic Waveguide

Abid Anjum Sifat,[1,2, a)] Ayed Al Sayem,[1,2] and M Mahmudul Hasan Sajeeb [1,3]

[1]*Department of EEE, Bangladesh University of Engineering and Technology, Dhaka, Bangladesh*

[2]*Department of EEE, Daffodil International University, Dhaka, Bangladesh*

[3]*Department of EEE, Ahsanullah University of Science & Technology, Dhaka, Bangladesh*

In this article, a 2D plasmonic waveguide loaded with all dielectric anisotropic metamaterial, consisting of alternative layers of Si-SiO$_2$, has been theoretically proposed and numerically analyzed. Main characteristics of waveguide i.e. propagation constant, propagation length and normalized mode area have been calculated for different values of ridge width and height at telecommunication wavelength. The respective 1D structure of the waveguide has been analytically solved for the anisotropic ridge as a single uniaxial medium with dielectric tensor defined by Effective Medium Theory (EMT). The 2D structure has been analyzed numerically through FEM simulation using Mode analysis module in Comsol Multiphysics. Both the EMT and real multilayer structure have been considered in numerical simulations. Such structure with all dielectric metamaterial provides an extra degree of freedom namely fill factor, fraction of Si layer in a Si-SiO$_2$ unit cell, to tune the propagation characteristics compared to the conventional DLSSP waveguide. A wide range of variations in all the characteristics have been observed for different fill factor values. Besides, the effect of the first interface layer has also been considered. Though all dielectric metamaterial has already been utilized in photonic waveguide as cladding, the implementation in plasmonic waveguide hasn't been investigated yet to our best knowledge. The proposed device might be a potential in deep sub-wavelength optics, PIC and optoelectronics.

**I. INTRODUCTION**

Optical waveguide is one of the substantial elements for on chip optical communication in photonic integrated circuit (PIC) technology [1-5]. The never-ending search of deeply sub-wavelength photonic devices for greater compactness, higher speed and reduced optical power has caused the emergence of optical components based on Surface Plasmon Polaritons (SPPs). Surface plasmon polaritons are surface waves originated from the collective oscillations of free electrons and photons which propagate along the interface of two media where real part of permittivity possesses opposite signs [5-10]. A variety of configurations such as Metal-Insulator-Metal (MIM) [11], Insulator-Metal-Insulator (IMI) [12], Dielectric loaded SPP (DLSPP) [4, 13-16], Channel Plasmon Polariton (CPP) [17, 18], Wedge SPP [19, 20], Hybrid Plasmon polariton (HPP) [21, 22], plasmonic waveguide cladded by hyperbolic metamaterial [23-25] etc. have already been proposed. But the incorporation of all dielectric metamaterials in plasmonic waveguides has not yet been explored to the best of our knowledge. After intensive investigation on metal and graphene based metamaterials [23-28], recently all dielectric metamaterial (ADM) [29] has emerged as a new metamaterial platform due to its lossless characteristics [30-33]. ADM is an artificially engineered structure, consisting of only dielectric materials, with optical properties which are not found in nature [30]. This overcomes the significant power loss in conventional metamaterials due to the presence of metal [30, 34]. ADMs have already shown exciting phenomenon such as skin depth engineering in photonic waveguides [31, 32, 35], broad angle negative refraction [36], dyakonov waves [37], bending loss reduction in SOI based waveguides [38], optical wave front molding [33, 39] etc. Dielectric nanostructures and all dielectric metasurfaces have also been found to provide spectrally selective electric and magnetic hotspots [40] for bio-sensing and energy applications [41, 42].

In this study, we have explored the potentials of ADMs constructed by multilayer of deeply sub-wavelength unit cells, consisting of two different dielectric materials, in a plasmonic waveguide. The ADM ridge, consisting of alternate layers of Si and $SiO_2$, has been placed on top of a thin gold metal film. Propagation characteristics of the respective 1D structure have been calculated analytically. The 2D structure of the proposed waveguide has been analyzed numerically by FEM (Finite Element Method) simulation in Comsol Multiphysics.

Important waveguide characteristics namely the propagation constant, propagation length and normalized mode area have been determined as a function of ridge width and height. In simulation, we have considered the real multilayer structure of the ADM ridge and also treated the multilayer structure as a uniaxial medium with dielectric tensor values determined by Effective Medium Theory (EMT). The extra degree of freedom of the proposed waveguide comes from the ratio of Si and $SiO_2$ in each unit cell which is known as the fill factor. With a fixed width and height, the whole waveguide performance can be modified solely by varying the fill factor. This is not possible for a conventional dielectric loaded plasmonic waveguide without changing the ridge material. As the fill factor is increased, the propagation constant follows a rising pattern whereas the propagation length gradually decreases. The field confinement has been found to improve with lower value of normalized mode area for larger fill factors. We have also considered the effect of first interface layer and slightly different waveguide characteristics have been found for the two different interface materials. Our proposed waveguide should be easily fabricated by current fabrication technology [43, 44] and may find suitable applications in deep sub-wavelength optics such as light guiding, routing, sensing etc.

## II. THEORY

The schematic of the proposed structure is shown in Fig. 1. The structure consists of a multilayer ridge with thickness $t_r$ and width w, on a thin metal (gold) film of thickness, $t_m$ and a semi-infinite glass substrate. The multilayer ridge is constructed by alternate stacking of Si and $SiO_2$. The thickness of the metal has been taken to be 100nm. The permittivity of the metal and substrate has been taken as $\varepsilon_3$= -86.64+8.74i and $\varepsilon_4$=2.56 [45] respectively. The anisotropic ridge is surrounded by air with permittivity $\varepsilon_1$=1.

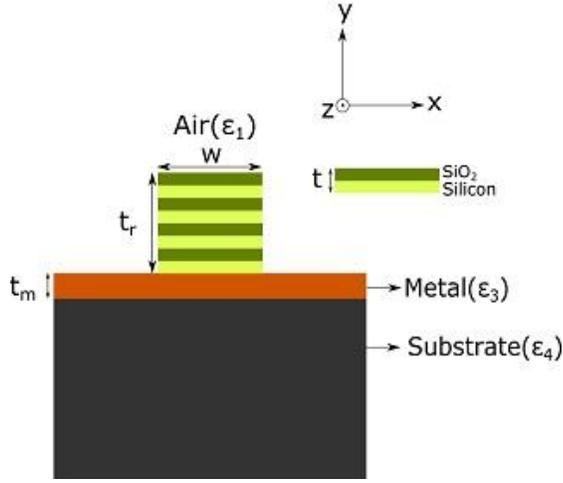

FIG. 1. Schematic of the proposed waveguide with Si-SiO$_2$ multilayer all dielectric metamaterial ridge of total thickness, $t_r$ and width, w on a thin metal (gold) film of thickness $t_m$ (100nm) and semi-infinite glass substrate. t is the thickness of the unit cell, $t_{Si}$ and $t_{SiO2}$ are the thicknesses of Si and SiO$_2$ respectively.

When the unit cell of the ADM ridge is deeply sub-wavelength, $k_o t \ll 1$ where $k_o = \frac{2\pi}{\lambda}$; $\lambda$ is the wavelength, the stack of alternate layers of Si and SiO$_2$ can be considered as an anisotropic metamaterial having the effective permittivity tensor as follows,

$$\varepsilon = \begin{bmatrix} \varepsilon_\parallel & 0 & 0 \\ 0 & \varepsilon_\perp & 0 \\ 0 & 0 & \varepsilon_\parallel \end{bmatrix} \tag{1}$$

Here, according to effective medium theory (EMT) [23- 25],

$$\varepsilon_\parallel = f\varepsilon_{Si} + (1-f)\varepsilon_{SiO_2} \tag{2a}$$

$$\varepsilon_\perp = \frac{1}{\frac{f}{\varepsilon_{Si}} + \frac{(1-f)}{\varepsilon_{SiO_2}}} \tag{2b}$$

$f$ is the fill factor defined as $f = \frac{t_{Si}}{t_{Si} + t_{SiO_2}}$.

The respective 1D structure (width, w=∞) of our proposed waveguide has been solved analytically applying EMT theory for TM polarized light and λ=1.55 μm [13, 21]. The dispersion relation has been derived as,

$$\varepsilon_\| k_{y2}(\varepsilon_3 k_{y3}(\varepsilon_4 k_{y1} + k_{y4}) + (\varepsilon_3{}^2 k_{y1} k_{y4} - \varepsilon_4 k_{y3}{}^2)\tan(t_m k_o k_{y3})) - (\varepsilon_3 k_{y3}(-\varepsilon_\|{}^2 k_{y1} k_{y4} + \varepsilon_4 k_{y2}{}^2) +$$
$$(\varepsilon_4 \varepsilon_\|{}^2 k_{y3}{}^2 k_{y1} + \varepsilon_3{}^2 k_{y4} k_{y2}{}^2)\tan(t_m k_o k_{y3}))\tan(t_r k_o k_{y2}) = 0 \qquad (3)$$

$k_{y4} = \sqrt{\beta^2 - \varepsilon_4}$, $k_{y3} = \sqrt{\varepsilon_3 - \beta^2}$, $k_{y2} = \sqrt{\varepsilon_\| - \frac{\varepsilon_\| \beta^2}{\varepsilon_\perp}}$, $k_{y1} = \sqrt{\beta^2 - 1}$, $k_o = \frac{2\pi}{\lambda}$ where, β is the propagation constant.

### III. RESULTS AND DISCUSSION

Fig. 2(a) shows the effective in plane and perpendicular permittivity of the multilayer anisotropic ADM as a function of fill factor. It can be observed that the effective parallel and perpendicular permittivity of the structure can be tuned between the values of the two constituent materials Si and SiO$_2$. Fig. 2(b) shows the propagation constant, β as a function of height of the 1D ADM ridge for TM0 (solid curve) and TM1 (dotted curve) mode for different fill factors. With the increase of the height $t_r$, β increases and eventually introduces TM1 mode at a height, $t_r$ ~250nm for fill factor 0.75. For lower values of fill factors, the TM1 mode originates at higher heights. This feature allows exciting only a single mode just by restricting the height of the ridge. The lower values of β with decreasing fill factors can be attributed to the fact that the high permittivity Si content decreases with the decrease of fill factor. This renders the in plane permittivity of the ADM to have smaller value according to (2a).

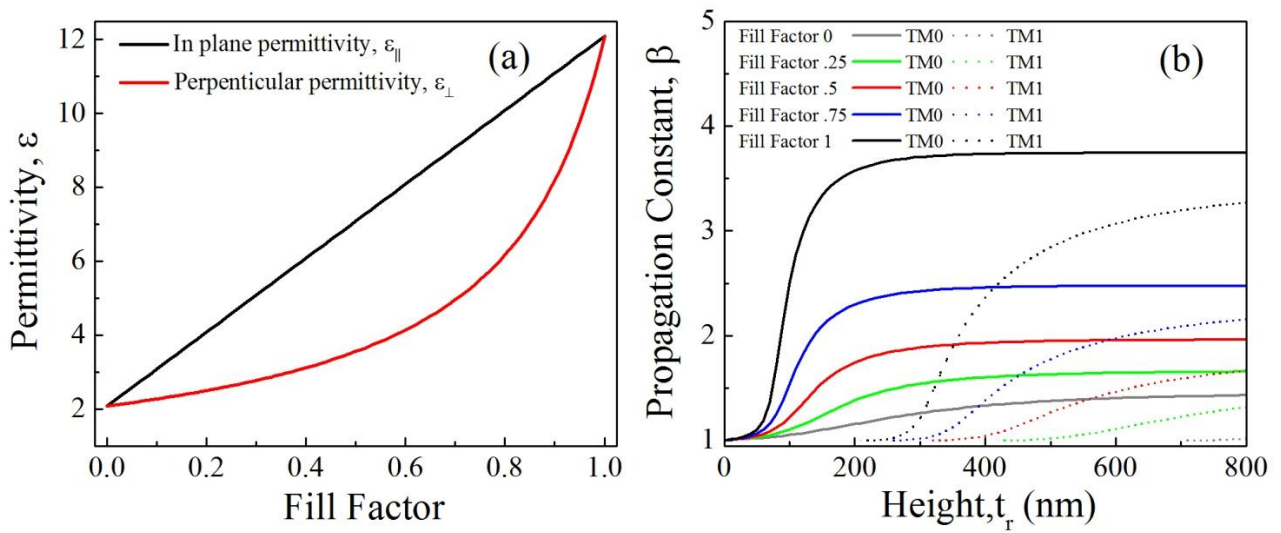

FIG. 2. (a) In plane permittivity, $\varepsilon_\|$ and perpendicular permittivity, $\varepsilon_\perp$ as a function of fill factor of the ADM ridge. (b) Propagation constant, β as a function of height of the semi-infinite (w=∞) ADM ridge for different fill factors. Solid and dotted lines represent the values for TM0 and TM1 modes respectively.

In order to find the propagation characteristics of the 2D structure (the width, w of the ADM ridge being finite) of the proposed device, Mode analysis module available in Comsol Multiphysics has been used. In simulation, we have considered both the real multilayer structure and the EMT.

EMT is basically derived for infinite periodic layers and layer order doesn't affect the result of EMT. But for practical finite structures with finite number of layers, layer order does play a significant role in device performance. In our simulation, we have considered both EMT and real multilayer structure made from alternate layers of Si and $SiO_2$ with two different layer ordering: i) Si as the first interface layer (Si on the top of Gold and then $SiO_2$ and repetition of this unit cell configuration) ii) $SiO_2$ as the interface layer ($SiO_2$ on the top of Gold and then Si and repetition of this unit cell configuration).

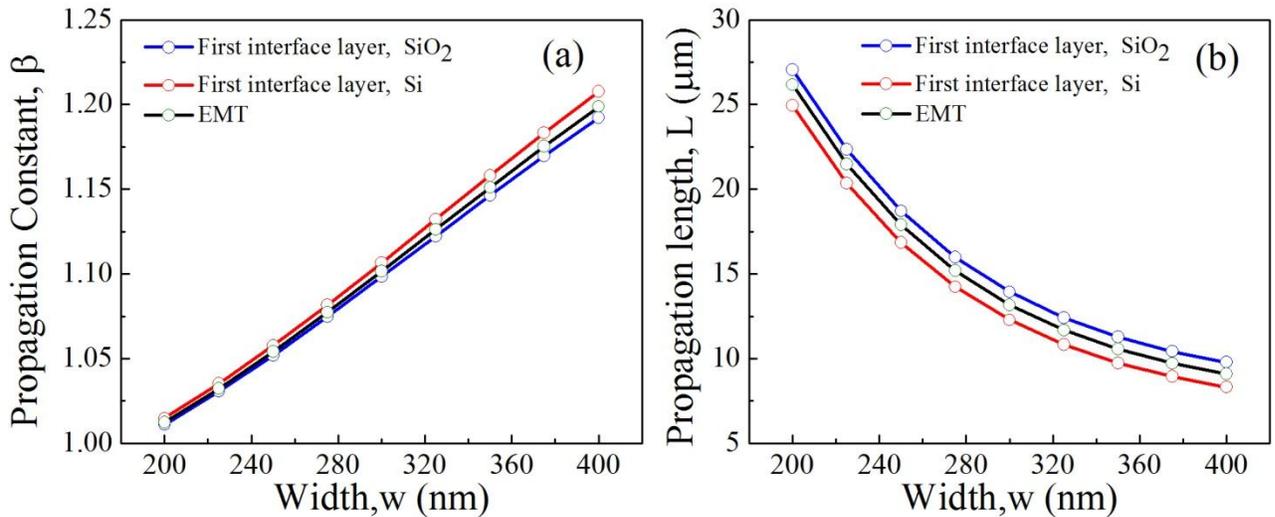

FIG. 3. (a) Propagation constant, β and (b) propagation length, L for the proposed 2D waveguide as a function of width, w of the 2D ADM ridge at a fixed height, $t_r$= 160 nm and fill factor f=0.5. Each unit cell has been taken to be of 20nm and total number of unit cell is eight.

Fig. 3(a) shows the propagation constant, β as a function of the 2D ADM ridge width, w at a fixed height, $t_r$= 160nm and fill factor f=0.5. Eight unit cells have been used where the thickness of each unit cell has been taken to be 20nm. It can be observed that with the increase of ridge width, the propagation constant increases almost linearly. The propagation constant is slightly higher for the case of i) where Si is the first interface layer than ii) where $SiO_2$ is the first interface layer. The value of propagation constant remains in between the values observed for the case i) and ii) when the multilayer ridge is replaced by a single medium with the permittivity tensor values calculated from EMT.

Propagation length is one of the most important characteristics of any waveguide which has been calculated as, $L = \frac{1}{2k_o Im(\beta)}$. Fig. 3(b) shows the propagation length, L as a function of width, w of the proposed structure at a fixed height, $t_r$= 160 nm and fill factor f=0.5. From Fig. 3(b), it can be observed that, propagation length, L decreases with the increase of width and the values obtained for EMT can be found in between the values acquired for the two cases (i and ii as mentioned above) of real multilayer structure. The propagation length has been observed to be comparable with the values found in literature for wave guiding application at telecommunication wavelength [5, 10].

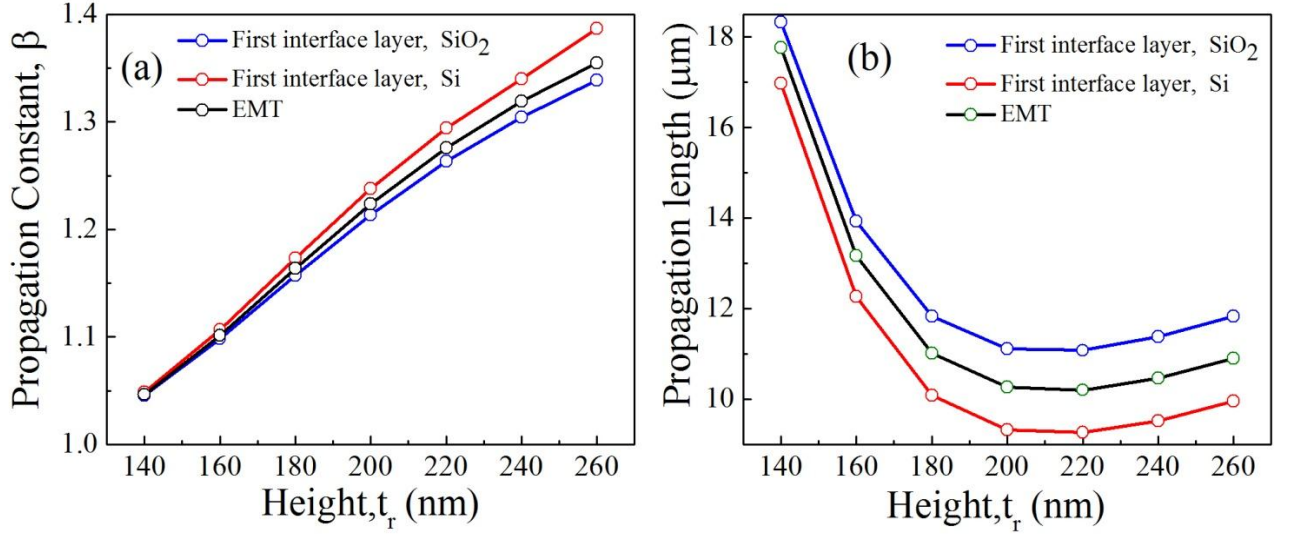

FIG. 4. (a) Propagation constant, β and (b) propagation length, L as a function of height, $t_r$ of the 2D ADM ridge at a fixed width, w=300nm and fill factor, f=0.5. All other parameters are the same as mentioned in Fig. 3.

The waveguide characteristics with the variation of height (number of unit cells) at fixed width have also been analyzed. Fig. 4(a) shows the propagation constant, β as a function of height, $t_r$ of the proposed waveguide at a fixed width, w=300nm and fill factor 0.5. It can be observed from the figure that β increases almost linearly with heights below 180-200 nm but the increment tappers as the height increases above 200nm. Also the propagation constant obtained for EMT remains in between the ones observed for the cases i) and ii) in real multilayer structure. Fig. 4(b) shows the propagation length, L as a function of height, $t_r$ at the same fixed width, w=300nm and fill factor 0.5. Propagation length, L shows a very interesting behavior with the variation of height of the ADM ridge. At a certain height (~210nm), the propagation length, L reaches a minimum for both cases i) and ii) and also for the EMT. Like the width variation, with the variation of height both the β and L values found for EMT lies in between the values obtained for the two cases (i and ii) of real multilayer structure.

Another important performance criterion of waveguide namely the normalized mode area, $A_m/A_o$ [16, 20, 46] has been calculated for the proposed structure as a function of both ridge height and width. Here, $A_o$ is the diffraction limited mode area in vacuum, $A_o=\lambda^2/4$. Mode area, $A_m$ is defined as the ratio of total mode energy density to maximum energy density [16, 20, 46] which can be given by,

$$A_m = \frac{\iint_{-\infty}^{\infty} W(r)\, d^2r}{\max(W(r))} \qquad (4)$$

where W(r) is the electromagnetic energy density (energy per unit volume) given by, $W(r) = \frac{1}{2}\left(\frac{d(\varepsilon(r)\,\omega)}{d\omega}|E(r)|^2 + \mu_o|H(r)|^2\right)$

Here, E(r) and H(r) are the electric field and magnetic field distribution in the structure.

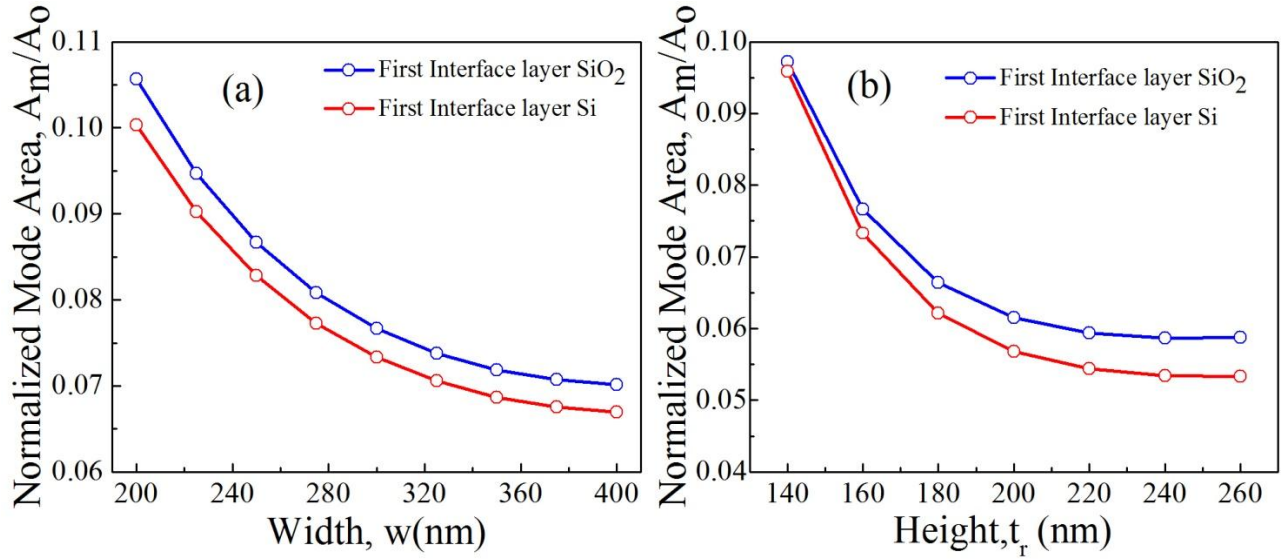

FIG. 5. Normalized mode area $A_m/A_o$ (a) as a function of ADM ridge width, w for a fixed height, $t_r$=160nm; (b) as a function of ADM ridge height, $t_r$ for a fixed width, w=300nm. In both cases the fill factor has been chosen to be 0.5. All other parameters are the same as mentioned in Fig. 3.

Fig. 5(a) and 5(b) shows the normalized mode area $A_m/A_o$ as a function of ADM ridge width at fixed height, $t_r$=160nm and as a function of ADM ridge height at fixed width, w=300nm respectively for the both cases i) and ii) of real multilayer structure as stated above. Here the fill factor of the ADM ridge has been fixed to 0.5. It can be observed from the Fig. 5(a) that the mode confinement increases with the increase of the width. This can be attributed to the higher index values associated with these higher widths shown in Fig. 3(a) [15, 16]. Normalized mode area for i) Si as the first interface layer is slightly lower than that for ii) $SiO_2$ as the first interface layer. This might be due to the higher propagation constant obtained for i) Si as the first interface layer than the other one. From Fig. 5(b), it can be observed that Normalized mode area decreases dramatically with the increase of height and then gets stabilized for heights larger than 240 nm for both cases i) and ii) of the real multilayer structure as stated above. And the trend obtained here can also be attributed to the higher index values associated with the higher heights.

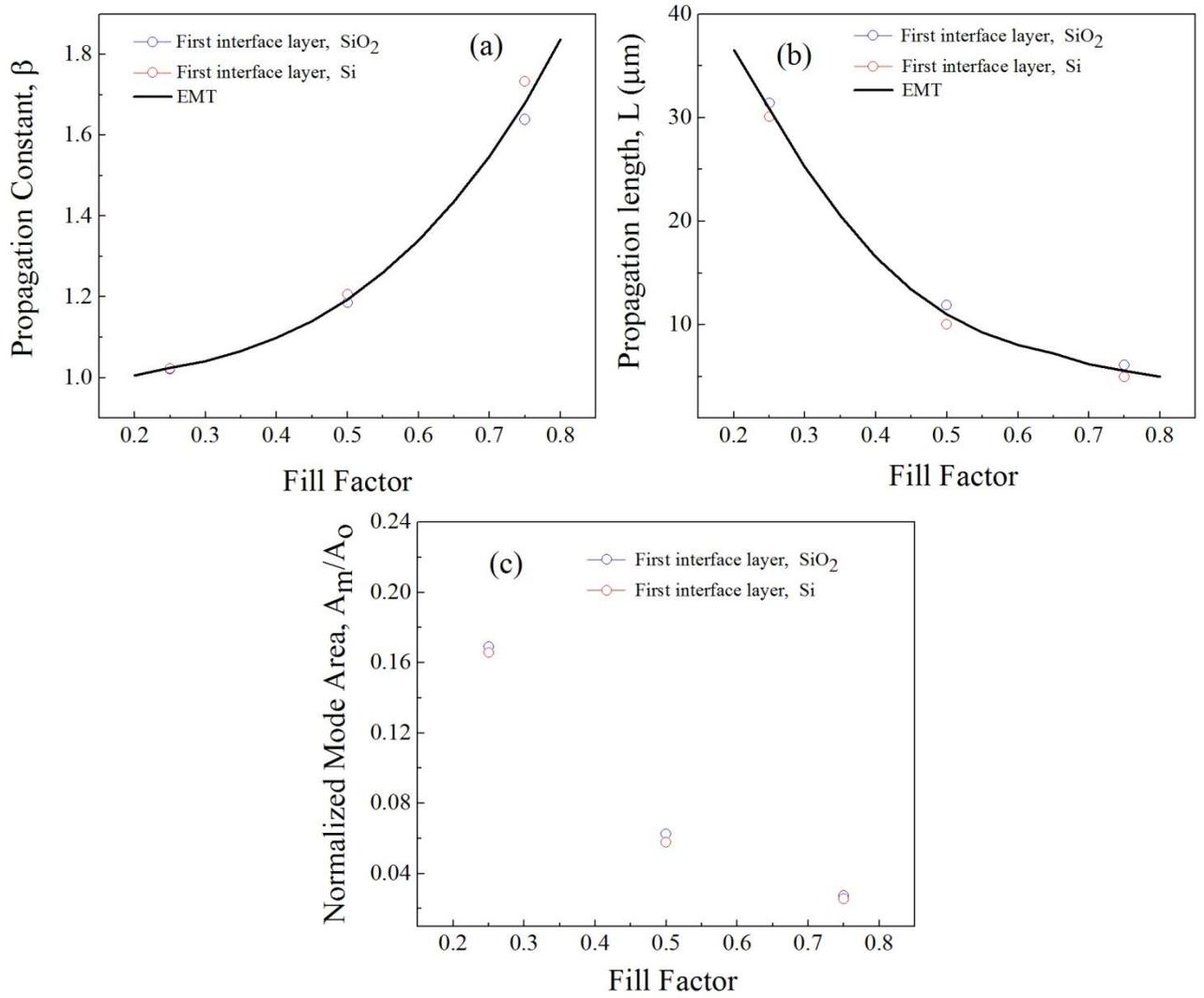

FIG. 6. (a) Propagation constant, β (b) propagation length, L and (c) normalized mode area, $A_m/A_o$ as a function of fill factor of the ADM ridge at a fixed width, w= 280nm and height, $t_r$=200nm. All other parameters are the same as mentioned in Fig. 3.

Finally the effect of fill factor on waveguide performance has been analyzed. Fig. 6(a), 6(b) and 6(c) show propagation constant, propagation length and normalized mode area respectively as a function of fill factor at a fixed width, w=280 nm and height, $t_r$=200 nm. From Fig. 6(a) it can be observed that the propagation constant, β increases with the increase of fill factor. This is expected as with the increase of fill factor, the effective parallel permittivity of the ADM ridge increases which can be observed from Fig. 2(a). Propagation length, L decreases with the increase of fill factor as evident from Fig. 6(b). Both EMT and real multilayer structure show very similar phenomenon. The β and L values, found for EMT, lie in between the values observed for case i) and case ii) of the multilayer structure as stated above. Finally from Fig. 6(c), it can be observed that with the increase of fill factor, the normalized mode area decreases indicating higher confinement which is also logical because of high in plane permittivity for higher fill factors of ADM ridge. The whole device characteristics can be modified by tuning the fill factor of the structure without changing the width and height of the structure. Such modification is not possible for conventional DLSPP waveguides [4, 13-16] without changing the ridge material. By using ADM and tuning the fill factor any desired effective permittivity valued ridge can be constructed and as a consequence desired waveguide performance can be attained.

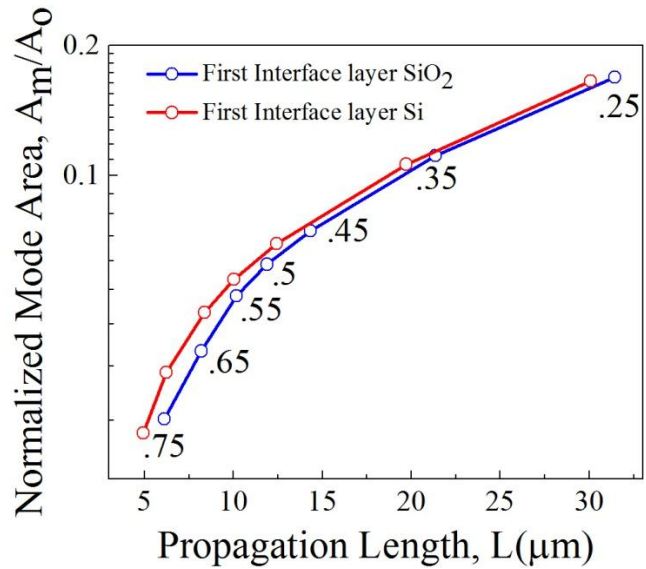

FIG. 7. (a) Normalized mode area, $A_m/A_o$ vs propagation length, L for different fill factors, the ADM ridge has a fixed width, w= 280nm and height, $t_r$=200nm. All other parameters are the same as mentioned in Fig. 3.

To have a much deeper insight of the effect of fill factor on waveguide characteristics, normalized mode area, $A_m/A_o$ found for the two cases i) and ii) of real multilayer structure is plotted with respect to propagation length, L for the variation of fill factors in Fig. 7. It can be clearly observed from the figure that with the increase of fill factors, propagation length and normalized mode area follow different trends; propagation length decreases and normalized mode area increases. So, the whole device performance can be modified with the variation of fill factors without changing the width and height of the waveguide which is not possible by any conventional plasmonic waveguides.

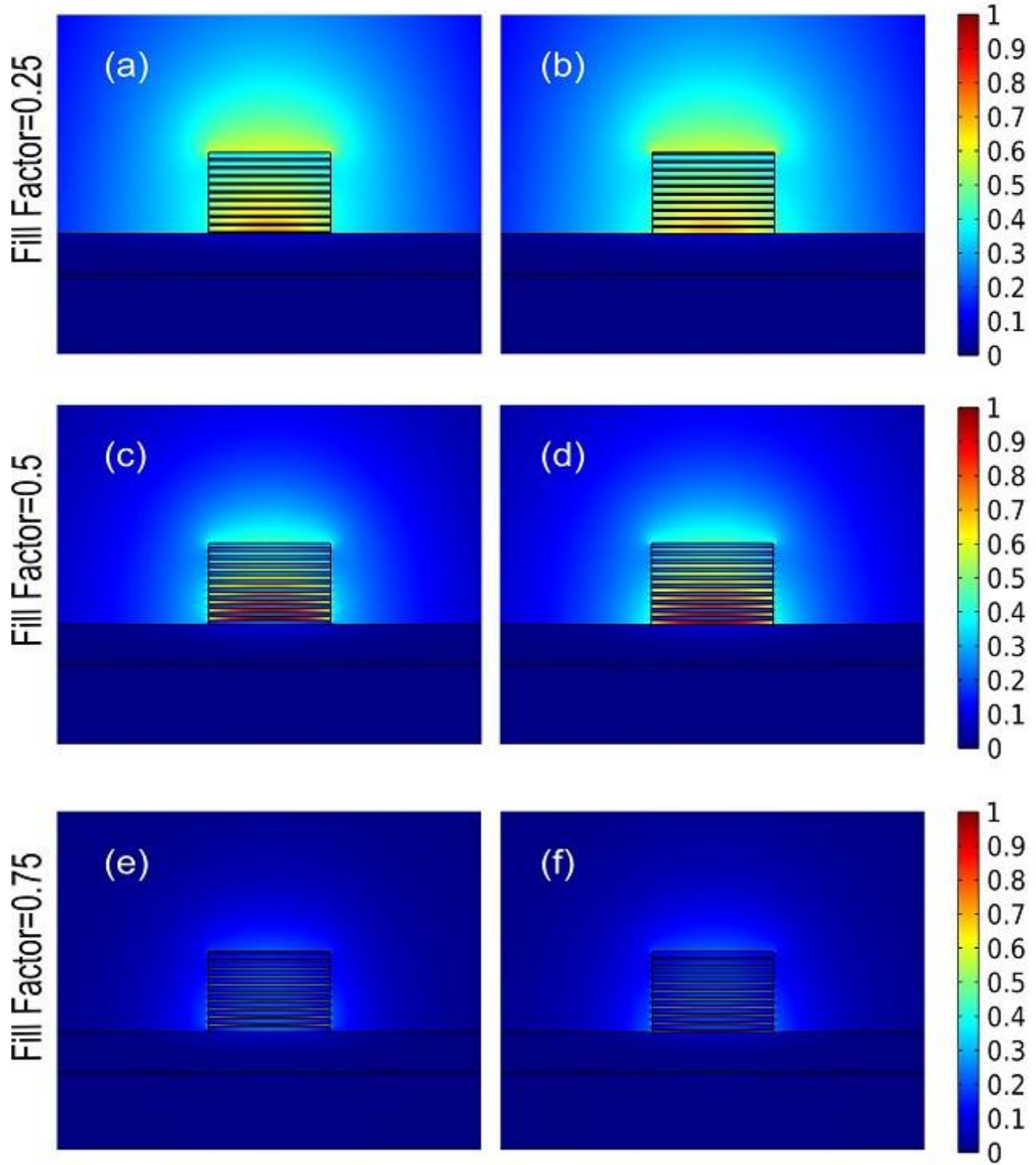

FIG. 8. Normalized electric field profile in the real multilayer structure of ADM ridge with a fixed height, $t_r$=200nm and width, w=300nm at a fill factor of (a, b) 0.25, (c, d) 0.5 and (e, f) 0.75. For (a, c and e) Si is the first interface layer and for (b, d and f) $SiO_2$ is the first interface layer.

Fig. 8 shows the normalized electric field profile in the real multilayer structure at a fixed height, $t_r$=200nm and width, w=300nm for three different fill factors 0.25, 0.5 and 0.75 considering both the cases i) Si as the first interface layer and case ii) $SiO_2$ as the first interface layer. It is clearly evident from Fig. 8 that the field confinement increases as the fill factor of the ADM ridge is increased, which has also been observed in Fig. 6(c). Besides, the field profile can be found to be almost similar for all the fill factors considered for both case i) and ii).

**IV. CONCLUSION**

In this article, a new type of 2D plasmonic waveguide has been proposed where an ADM ridge is placed on top of a thin metal film. ADM consists of multilayer of deeply sub-wavelength loss less dielectric/semiconductor materials Si and $SiO_2$. Optimally chosen ridge structure can provide sufficient propagation length with deeply sub-wavelength confinement. The whole waveguide performance can be tuned by varying the fill factor of the metamaterial keeping both the width and height of the ridge constant which is not possible by conventional dielectric loaded plasmonic waveguide. Both real multilayer structure and EMT show very similar phenomena. The proposed device can be easily fabricated by current fabrication technology and also the most common materials have been used in this article. The proposed device may find fundamental applications in photonic integrated circuits, wave-guiding, sensing etc.